\documentclass[nonacm,sigconf]{acmart}

\usepackage{adjustbox}
\usepackage{float}
\usepackage{comment}
\usepackage{multirow}
\usepackage{array}
\usepackage{xspace}
\usepackage{color}
\usepackage{caption}
\usepackage{tabularx}
\usepackage{array}
\usepackage{ragged2e}
\usepackage{subcaption}
\usepackage{url}
\usepackage{graphicx}
\usepackage{enumitem}
\newcommand{\methodName}{RAM\xspace}

\setcopyright{none}
\renewcommand\footnotetextcopyrightpermission[1]{} 

\begin{document}
%
\title{Rule-ATT\&CK Mapper (\methodName): Mapping SIEM Rules to TTPs Using LLMs}

\author{Prasanna N. Wudali}
\affiliation{%
  \institution{Ben-Gurion University of the Negev}
  \country{}
}

\author{Moshe Kravchik}
\affiliation{%
  \institution{Rafael Advanced Defense Systems}
  \country{}
}

\author{Ehud Malul, Parth A. Gandhi, Yuval Elovici, Asaf Shabtai}
\affiliation{%
  \institution{Ben-Gurion University of the Negev}
  \country{}
}
\renewcommand{\shortauthors}{}

\begin{abstract}
The growing frequency of cyberattacks has heightened the demand for accurate and efficient threat detection systems. 
Security information and event management (SIEM) platforms are important for analyzing log data and detecting adversarial activities through rule-based queries, also known as SIEM rules. 
The efficiency of the threat analysis process relies heavily on mapping these SIEM rules to the relevant attack techniques in the MITRE ATT\&CK framework. 
Inaccurate annotation of SIEM rules can result in the misinterpretation of attacks, increasing the likelihood that threats will be overlooked. 
Such misinterpretation can expose an organization's systems and networks to potential damage and security breaches.
Existing solutions for annotating SIEM rules with MITRE ATT\&CK technique and sub-technique labels have notable limitations: 
manual annotation of SIEM rules is both time-consuming and prone to errors, and machine learning-based approaches mainly focus on annotating unstructured free text sources (e.g., threat intelligence reports) rather than structured data like SIEM rules. 
Structured data often contains limited information, further complicating the annotation process and making it a challenging task.
To address these challenges, we propose Rule-ATT\&CK Mapper (\methodName), a novel framework that leverages large language models (LLMs) to automate the mapping of structured SIEM rules to MITRE ATT\&CK techniques. 
\methodName's multi-stage pipeline, which was inspired by the prompt chaining technique, enhances mapping accuracy without requiring LLM pretraining or fine-tuning. 
Using the Splunk Security Content dataset, we evaluate \methodName's performance using several LLMs, including GPT-4-Turbo, Qwen, IBM Granite, and Mistral. 
Our evaluation highlights GPT-4-Turbo’s superior performance, which derives from its enriched knowledge base, and an ablation study emphasizes the importance of external contextual knowledge in overcoming the limitations of LLMs' implicit knowledge for domain-specific tasks. 
These findings demonstrate \methodName's potential in automating cybersecurity workflows and provide valuable insights for future advancements in this field.

\end{abstract}


\maketitle

\thispagestyle{empty}

\section{\label{sec:intro}Introduction}

The rapid advancement of technology and widespread adoption of digital applications have resulted in a significant increase in cyberattacks~\cite{checkpoint}.
To gain visibility into their digital ecosystems, organizations deploy security information and event management (SIEM) systems in their networks. 
These systems store and analyze log data generated by various digital entities in the network~\cite{exabeam}.

SIEM systems enable threat detection by allowing users to execute search queries, referred to as rules, on the ingested log data. 
Each SIEM platform employs its own rule definition language (RDL), a schema-based structure for defining these rules that standardizes the creation and execution of SIEM rules, making them inherently structured data and a foundational component of modern cybersecurity operations.
Examples of such schemas include the search processing language (SPL) from Splunk, the Lucene query language by Elasticsearch, and the Kusto query language (KQL) by Microsoft. 

Security alerts are triggered when the execution of SIEM rules yields search results. 
When such alerts are generated, security analysts must examine each alert individually, performing tasks such as triage, analysis, and interpretation, and determine whether the alert corresponds to an actual attack. 
A critical aspect of effective threat detection and hunting is the precise mapping and understanding of the tactics, techniques, and procedures (TTPs) employed by adversaries, as defined in the MITRE ATT\&CK framework.\footnote{\url{https://attack.mitre.org/}} 
Incorporating MITRE ATT\&CK techniques in the analysis provides valuable insights, enabling analysts to discern potential attack flows. 
Such mapping enhances security professionals' ability to anticipate and mitigate the strategies employed by cyber adversaries.

Mapping SIEM rules to specific MITRE ATT\&CK techniques is a complex manual process that is prone to errors and can be time-consuming.
Cybero, a leading cybersecurity company, reported~\cite{cybero} that \textit{"organizations collect sufficient log data to potentially detect 94\% of techniques outlined in the MITRE ATT\&CK framework; however, only 24\% of these techniques are effectively covered due to gaps in detection rules, with an additional 12\% of SIEM rules rendered non-functional or misconfigured."} 
In its best practices guide~\cite{cisa} to MITRE ATT\&CK mapping, CISA, an American cyber defense agency, listed (i) leaping to conclusions (i.e., prematurely deciding on a mapping based on insufficient evidence or examination of the facts), (ii) missing opportunities (i.e., not considering, being unaware of, or overlooking other potential technique mappings based on implied or unclear information), and (iii) miscategorization (i.e., the selection of incorrect techniques due to misinterpreting, misreading, or inadequately understanding the techniques, specifically the difference between two techniques) as common mistakes committed by security analysts when manually performing the mapping task.
Given the above, there is a need to automate the mapping process and thereby reduce the workload on security analysts and increase the speed and accuracy of threat detection.

Recent cybersecurity research has explored various techniques for mapping unstructured data from cyber threat intelligence (CTI) reports to the MITRE ATT\&CK framework~\cite{alves2022leveraging,alam2023looking,rani2024ttpxhunter,liu2022threat,zhang2024attackgboosting}. 
While these methods have demonstrated effectiveness in handling unstructured data, they have a limited ability to adapt to structured data use cases, such as intrusion detection system and SIEM rules.
Also, these methods use supervised learning-based approaches to classify structured data (i.e., intrusion detection system and SIEM rules) to MITRE ATT\&CK technique classes, which require retraining when new threats emerge.
Their reliance on retraining limits their scalability and efficiency in dynamic threat landscapes.
Mărmureanu et al.~\cite{10398612} proposed a method to map structured data, specifically Splunk rules, to the MITRE ATT\&CK framework. 
This approach utilizes a BERT model trained as a classifier to categorize Splunk rules into 14 high-level MITRE ATT\&CK tactic classes. 
However, this method shares the same limitations as other supervised learning approaches discussed earlier, particularly the need for retraining with updated data to address new threats. 
Furthermore, the task of mapping rules to high-level tactics is comparatively easier than mapping them to MITRE ATT\&CK techniques and sub-techniques, which involve around 670 distinct classes and present a much greater challenge. 
Despite focusing on this simplified task, the method failed to achieve high performance in their evaluation, due to its inherent limitations.
In a recent study, Fayyazi et al.~\cite{fayyazi2023advancing} employed large language models (LLMs) to map CTIs in the form of unstructured text to MITRE ATT\&CK techniques, while Nir et al.~\cite{daniel2023labeling} employed them to map Snort intrusion detection rules to MITRE ATT\&CK techniques. 

These investigations highlight the potential of LLMs in cybersecurity tasks but also underscore their limitations. 
Solely relying on the implicit knowledge of LLMs has proven insufficient for addressing the domain-specific requirements of cybersecurity.
This gap highlights the need for more adaptable and scalable methodologies tailored to the dynamic nature of cyber threats.
To produce accurate and reliable predictions, they require additional contextual information that is not inherently available to the LLM.

To address these shortcomings, we propose \methodName, a novel LLM-based framework for analyzing SIEM rules and recommending relevant MITRE ATT\&CK techniques. 
\methodName eliminates dependence on training data, utilizes LLM agents to retrieve supplementary contextual information, and transforms structured rule into unstructured natural language to preserve the syntactic and semantic meaning of the rule.
This innovative approach ensures reliable and accurate predictions while overcoming the limitations of existing methods.

LLMs, with their advanced natural language processing (NLP) capabilities, can process and analyze structured data, automatically identify patterns, and understand the syntactic meaning of the data, but they often fall short in understanding the semantic meaning of the data.
This study leverages LLMs to autonomously map structured data in the form of SIEM rules 
to MITRE ATT\&CK techniques, enabling the automation of cybersecurity threat detection and investigation.

\begin{figure*}[ht!]
    \centering
    \includegraphics[width=0.85\linewidth]{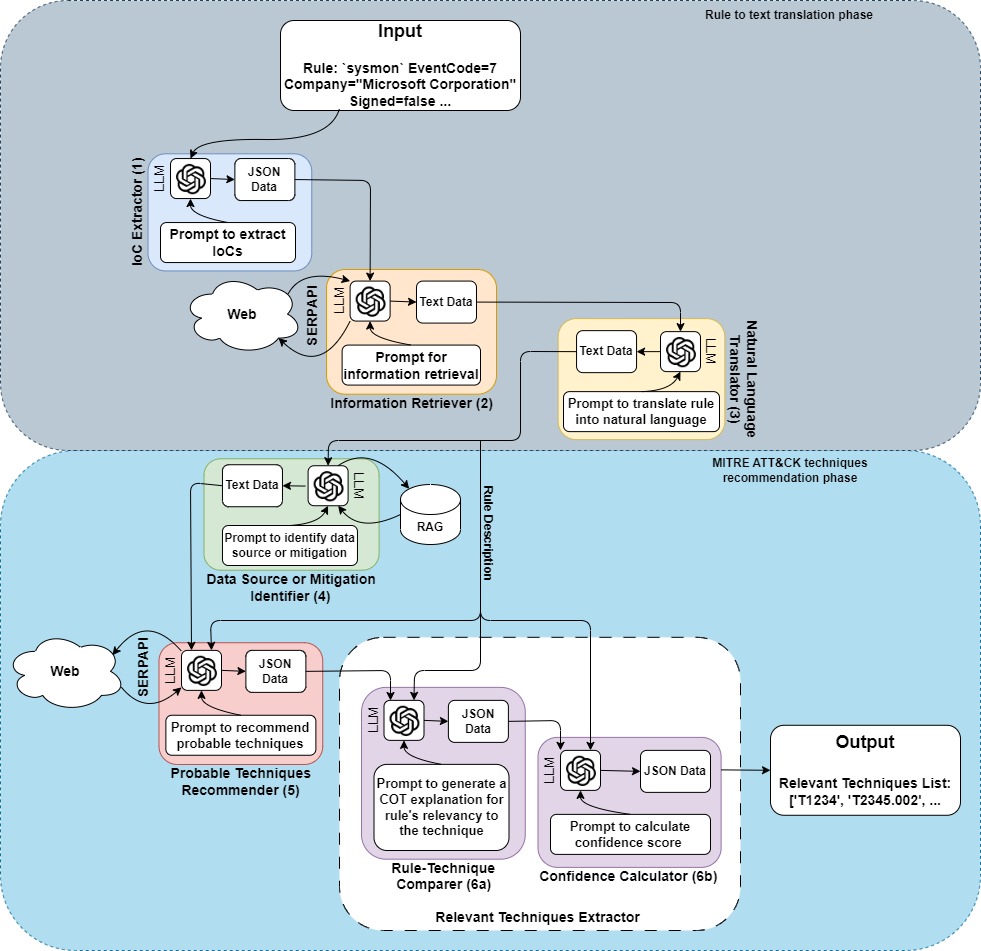}
    \caption{Overview of our AI Agent-based \methodName pipeline.}
    \label{fig:r2t}
\end{figure*}

\methodName is a multi-stage AI agent pipeline (see Figure~\ref{fig:r2t}) inspired by the prompt chaining technique~\cite{wu2022promptchainer} and designed to enhance the understanding and application of SIEM rules. 
The pipeline begins with the extraction of indicators of compromise (IoCs) from the rule (e.g., process names, file names, registry keys and values, IP addresses, network ports).
Then, a web search LLM agent retrieves additional contextual information related to the IoCs identified in the rule. 
Leveraging the information gathered in the preceding stages, the next AI agent translates the rule into natural language text, providing a comprehensive description.
This textual description is then used by an LLM to identify the data source~\cite{ds} of the logs or the mitigation strategy being applied upon which the rule operates.
This natural language representation, along with the data source or mitigation-related information, serves as input to another LLM that maps the rule in question to probable MITRE ATT\&CK techniques. 
In the final stage, the pipeline refines the mapping and provides reasoning by extracting the most relevant techniques from the list of potential matches, facilitating precise alignment of the rule with the MITRE ATT\&CK framework.

We conducted a comprehensive series of experiments to evaluate \methodName's ability to map SIEM rules to the MITRE ATT\&CK framework. 
The evaluation focused on common metrics such as precision and recall, which are indicators of the method's accuracy and completeness in correctly classifying the SIEM rules to relevant techniques within the framework.
Various LLMs were examined, including Qwen, IBM Granite, Mistral, and GPT-4-Turbo, and we evaluated \methodName's effectiveness when each LLM was employed in the pipeline.
We used the threat detection rules published in the Splunk Security content dataset\footnote{\url{https://github.com/splunk/security_content/tree/develop/detections}} in our experiments; to ensure that the rules were not already known to the LLM, we carefully selected rules for the dataset based on their creation or modification dates. 
Specifically, we included only those rules with dates later than the knowledge cut-off date of the LLMs utilized in our experiments.

Using various configuration settings, we aimed to identify the optimal strategies for maximizing the performance of these models. 
Our study not only demonstrates the potential of LLMs in automating threat analysis but also provides insights into the most effective configurations for deploying these models in real-world cybersecurity environments. 
This study is among the first to explore leveraging LLMs to map structured data to the MITRE ATT\&CK framework, and our results, which highlight RAM's potential, leave room for further refinement in future research. 
We also provide valuable insights regarding the challenges encountered during this study, which can guide subsequent advancements in this domain for example, the lack of a completely labeled SIEM rules dataset.

\noindent The main contributions of this paper are as follows:

\begin{itemize}[nosep,leftmargin=*]
    \item We demonstrate the feasibility of using LLMs to automate the mapping of SIEM rules to MITRE ATT\&CK techniques and provide reasoning, which could significantly enhance the capabilities of current cybersecurity tools.
    \item We propose an AI agent-based framework that utilizes both implicit and explicit knowledge in automating the mapping of structured SIEM rules to MITRE ATT\&CK techniques.
    \item We demonstrate the effective utilization of LLMs without the need for pretraining or fine-tuning, thereby eliminating the need for any training data.
    \item We provide a practical guide for deploying LLMs in cybersecurity, by identifying the optimal configurations for these models.
   
    \item We present valuable insights regarding the challenges encountered during the experimentation process, providing increased understanding of the obstacles and considerations that shaped our research and findings.
\end{itemize}

\section{\label{sec:background}Background}

LLMs, due to their advanced NLP and generation capabilities, are well-suited to analyze structured data by understanding its syntactical meaning, as well as for providing human understandable reasoning. 
However, effectively mapping structured data, such as SIEM rules, to the MITRE ATT\&CK framework~\cite{al2024mitre} requires not only syntactical understanding but also a deep comprehension of the semantic meaning of the rules. 
LLMs must discern the semantic content of a rule and identify its alignment with the techniques and sub-techniques. 
Relying solely on the implicit knowledge of LLMs is insufficient for this complex task.

To address this challenge, various prompt engineering techniques~\cite{sahoo2024systematic} were implemented to enhance the LLMs' ability to understand the semantic nuances of SIEM rules and propose relevant MITRE ATT\&CK techniques and sub-techniques. 
These prompt engineering techniques are discussed in detail in this section.
Additionally, to provide context for readers unfamiliar with the MITRE ATT\&CK framework, a brief overview of its structure and purpose is also included in this section.

\subsection{MITRE ATT\&CK Framework}
The MITRE ATT\&CK framework~\cite{al2024mitre} is a comprehensive, globally recognized knowledge base that provides detailed insights into the tactics, techniques, and procedures (TTPs) used by adversaries in cyberattacks. 
It is designed to aid cybersecurity professionals in understanding adversarial behaviors, identifying attack patterns, and strengthening defensive strategies. 
By systematically categorizing adversarial actions, the framework helps organizations enhance their detection, response, and mitigation capabilities.

The framework is structured around three primary components: tactics, techniques, and procedures. 
\textit{Tactics} represent the high-level objectives that adversaries aim to achieve during an attack. 
These objectives outline the “why” behind an adversary’s actions and are categorized into stages of an attack lifecycle, such as Initial Access, Execution, Persistence, Privilege Escalation, and Exfiltration. 
\textit{Techniques} 
define the specific methods used to accomplish these tactical goals. 
These represent the “how” of an attack and include actions like phishing (to gain initial access), command and scripting interpreter usage (for execution), or credential dumping (to gain access to sensitive credentials). 
Complementing these, \textit{procedures} provide real-world examples of how techniques are operationalized, offering context on specific tools, scripts, or strategies used in documented adversarial campaigns.

In addition to TTPs, the MITRE ATT\&CK framework introduces two critical concepts: data sources and mitigations. 
\textit{Data sources} refer to the various types of telemetry and system-generated data that can be collected and analyzed to detect adversarial techniques.
\textit{Mitigations} describe the preventive or corrective actions that can be implemented to neutralize threats or limit their impact.

\subsection{Prompt Engineering}

Prompt engineering~\cite{sahoo2024systematic} is the practice of crafting precise and effective input prompts to optimize the performance and output of LLMs.
It involves designing queries or instructions that guide the model's understanding and execution of tasks. 
This technique has become pivotal in ensuring that LLMs generate accurate, contextually appropriate, and reliable results, especially in tasks requiring nuanced reasoning or complex problem-solving.

\subsubsection{Prompt Chaining}

Prompt chaining~\cite{wu2022promptchainer} is a technique that decomposes complex tasks into a sequence of smaller, logically ordered steps. 
In this approach, the output from one step serves as the input for the next, enabling a modular and iterative resolution of intricate problems. 
For instance, when generating a report, the initial prompt could request an outline, subsequent prompts could expand individual sections, and a final prompt could synthesize the results into a cohesive document. 
This method improves the clarity and manageability of multifaceted tasks.

\subsubsection{Chain-of-thought Prompting}
Another notable technique is chain-of-thought (CoT) prompting~\cite{wei2022chain}, which implements a step-by-step reasoning.
By explicitly including intermediate reasoning steps within the prompt or instructing the model to generate these steps, CoT prompting enhances the AI's capability to address tasks that require logical inference or multi-step computation.

\subsection{LLM Agents and React Framework}
LLM agents, or AI Agents~\cite{llmagents}, leverage the capabilities of large language models to perform tasks autonomously by reasoning, planning, and acting based on input instructions. 
They are typically used in applications like chatbots, decision-making systems, or task automation. 
These agents operate by combining LLMs with external tools, APIs, or environments to handle complex tasks that require more than natural language generation.

The REACT (REasoning \& ACTing) framework~\cite{yao2023reactsynergizingreasoningacting} enhances the functionality of LLM agents by combining logical reasoning with actionable operations in a unified system. 
REACT agents dynamically integrate high-level reasoning and decision-making with the execution of actions in an iterative feedback loop. 
This enables the agent to automatically analyze tasks, plan appropriate steps, execute actions, and adapt to new information or evolving contexts.
The REACT framework operates through a structured workflow: (1) The LLM interprets the input query or instruction, performs logical analysis, and generates a plan of action, (2) The agent executes the planned actions, such as querying an API or controlling an external system, (3) The results of the actions are analyzed, allowing the agent to refine its reasoning and plan subsequent steps, and (4) The final output integrates the outcomes of reasoning and acting, delivering a comprehensive response to the user.
\section{\label{sec:related}Related Work}

Table~\ref{tab:related_work} summarizes the related work on mapping cybersecurity data to attack tactics and techniques.
As can be seen, previous works mainly focused on mapping unstructured data, such as threat intelligence reports and semi-structured data, such as event logs, to MITRE ATT\&CK techniques. 
These efforts proposed both rule-based methods~\cite{kryukov2022mapping} and machine learning (ML)-based approaches~\cite{alam2023looking,alves2022leveraging,fayyazi2023advancing,rani2024ttpxhunter,you2022tim,liu2022threat,fengrui2024few,zhang2024attackgboosting}.

The rule-based method proposed by Kryukov et al.~\cite{kryukov2022mapping} aimed to map security events in SIEM to MITRE ATT\&CK framework using pre-defined rules based on threat patterns.
A key limitation of their method lies in its heavy reliability on rule (or patterns) database. 
While these patterns are essential to carry out accurate mapping, the method's reliance on them constrains its adaptability to newly emerging threats within the dynamic and ever-evolving cybersecurity landscape.
As a result, increased false positives (misidentification of benign activities as threats) and false negatives (failure to detect actual threats) are produced, particularly when new attack patterns or techniques have not been added to the database.

Recent advancements have marked a significant shift from traditional rule-based methods to the adoption of ML-based approaches, particularly the use of language models, for this task.
Language models, such as BERT and GPT, offer the ability to process unstructured text with minimal feature engineering due to their powerful contextual understanding and pre-trained embeddings.

You et al.~\cite{you2022tim} introduced a classification-based model for mapping unstructured data in the form of CTI reports to the MITRE ATT\&CK framework. 
Their approach utilized a combination of a bi-directional LSTM model and a CNN model to classify unstructured data into just six technique classes, significantly simplifying the task. 
However, the complexity of mapping unstructured data increases substantially when the model must predict across the entire set of technique classes in the MITRE ATT\&CK framework, which includes approximately 670 classes. 
Additionally, it is important to note that an attack pattern or rule can be mapped to multiple techniques, and a single technique may fall under multiple tactics, further complicating the mapping process.
This limitation may restrict the method’s applicability in real-world scenarios where comprehensive coverage of TTP classes is essential.

Liu et al.~\cite{liu2022threat} proposed a novel approach to map unstructured CTI to MITRE ATT\&CK framework. 
Their methodology, referred to as ATHRNN (attention-based transformer hierarchical recurrent neural network), employs a two-step classification process: first, classifying the unstructured text into MITRE ATT\&CK tactics and then further classifying it into MITRE ATT\&CK techniques.
In other work, Alves et al.~\cite{alves2022leveraging} explored the application of BERT models for the classification of unstructured text into MITRE ATT\&CK techniques.
The study uses eleven different BERT models to map unstructured texts to the MITRE ATT\&CK framework, aiming to enhance automation in cyber threat intelligence.

Similarly, Alam et al.~\cite{alam2023looking} proposed LADDER, a framework designed to enhance cybersecurity by automatically extracting attack patterns from CTI sources.
LADDER uses different BERT-based models for extracting attack patterns from unstructured texts and then mapping these patterns to MITRE ATT\&CK framework.
Rani et al.~\cite{rani2024ttpxhunter} proposed TTPXHunter, a method designed for the automated mapping of attack patterns extracted from cyber threat reports to MITRE ATT\&CK framework. 
This method is an extension of TTPHunter~\cite{ttp_hunter}, improving its ability to cover a broader range of techniques from MITRE ATT\&CK framework and precision with the help of a cyber domain-specific language model called SecureBERT.
Sentences are transformed into embeddings using SecureBERT and then sent to a linear classifier for TTP prediction.

Fayyazi et al.~\cite{fayyazi2023advancing} evaluated how well LLMs, specifically encoder-only (e.g., RoBERTa) and decoder-only (e.g., GPT-3.5) models, can summarize and map cyberattack procedures to the appropriate ATT\&CK tactics.
The authors compared various mapping approaches. 
Howerver, they focused on mapping cyberattack procedures to MITRE ATT\&CK tactics (which represent higher-level categorizations in MITRE ATT\&CK framework). 
While this is useful, it is comparatively easier than mapping cyberattack procedures to MITRE ATT\&CK techniques and sub-techniques, which are more granular and detailed, offering deeper insights into the specific actions and methods employed in an attack.

Fengrui et al.~\cite{fengrui2024few} introduced a method combining data augmentation and instruction supervised fine-tuning using LLMs to classify TTPs effectively in scenarios with limited data.
Similarly, Zhang et al.~\cite{zhang2024attackgboosting} introduced a novel framework for constructing attack knowledge graphs (KGs) from CTI reports, by leveraging LLMs. 

While these methods demonstrate remarkable progress in handling unstructured data, their applicability to structured data use cases, such as mapping SIEM rules, is limited. 
These approaches are specifically designed for unstructured data, where relationships between entities and contextual information are often explicitly defined, simplifying the mapping process to the MITRE ATT\&CK framework. 
Adapting these methods to structured formats like SIEM rules would require extensive modifications, reducing their effectiveness and suitability for such scenarios.

To the best of our knowledge, only two studies have specifically focused on mapping structured data, such as IDS rules and SIEM rules, to the relevant MITRE ATT\&CK technique (or sub-techniques) using language models.

Nir et al.~\cite{daniel2023labeling} investigated the integration of LLMs, specifically ChatGPT, into cybersecurity workflows to automate the association of network intrusion detection system (NIDS) rules with corresponding MITRE ATT\&CK techniques. 
While their method represents one of the first applications of LLMs for this purpose, their findings highlight the necessity of incorporating additional contextual information to enhance the accuracy and reliability of LLM predictions.
Mărmureanu et al.~\cite{10398612} proposed method to map structured data in the form of Splunk rules to MITRE ATT\&CK framework.
The authors proposed the use of ML classifiers to map the Splunk rules to tactics specified in MITRE ATT\&CK framework.
A significant limitation of these methods is their dependence on supervised learning approaches to train the machine learning models within their frameworks. 
These models are unable to dynamically adapt to evolving threat landscapes or newly introduced MITRE ATT\&CK techniques without undergoing retraining with updated datasets. 
This retraining process is not only time-consuming but also resource-intensive, substantially restricting the methods' ability to keep pace with the rapid evolution of cyber threats.

In summary, the limitations of prior studies can be categorized as follows: (1) reliance on supervised learning tasks, (2) inability to adapt to structured texts, and (3) dependence on additional contextual information to effectively interpret rules. To address these shortcomings, we propose \methodName, a novel LLM-based approach that eliminates the need for training data, coherently processes structured text into natural language, and employs LLM agents to retrieve supplementary contextual information, enabling reliable and accurate predictions.

\begin{table*}[h]
\renewcommand{\arraystretch}{1.5} 
\centering
\footnotesize
\caption{Summary of previous work.}

\begin{tabular}{|>{\raggedright\arraybackslash}p{2cm}|>{\raggedright\arraybackslash}p{1.4cm}|>{\raggedright\arraybackslash}p{2.5cm}|>{\raggedright\arraybackslash}p{4.7cm}|>{\raggedright\arraybackslash}p{5.8cm}|}
\hline
\textbf{Paper name} & \textbf{Mapping} & \textbf{Input data (\& type)} & \textbf{Method} & \textbf{Comments}\\ \hline
Kryukov et al.~\cite{kryukov2022mapping} (2022) & Techniques & SIEM alerts log (semi-structured) & Rule-based mapping & Coverage of all technique classes. No quantitative metric was provided.\\ \hline

Alves et al.~\cite{alves2022leveraging} (2022) & Techniques & MITRE Procedures (unstructured) & Train BERT model as classifier & Achieved classification accuracy of 0.82 on test dataset. Coverage of only 253 techniques in their evaluation. \\ \hline

You et al.~\cite{you2022tim} (2022) & Techniques & CTI reports (unstructured) & Use pre-trained Sentence-BERT for embeddings; train bi-LSTM with attention coupled with CNN for classification & Evaluation performed on only six techniques which extremely simplifies the classification task. Classification accuracy on six techniques is 0.94. \\ \hline

Liu et al.~\cite{liu2022threat} (2022) & Techniques & CTI reports (unstructured) & Train transformer and RNN-based model as classifier & Coverage of all technique classes. Achieved an AUC score of 0.76 during the classification task.\\ \hline

Fayyazi et al.~\cite{fayyazi2023advancing} (2023) & Tactics & MITRE Procedures (unstructured) & RAG-based approach to improve LLM performance & Only 14 classes available for classification. Achived a high F1 score of 0.95 when using RAG to fetch external data.\\ \hline

Alam et al.~\cite{alam2023looking} (2023) & Techniques & CTI reports (unstructured) & Train BERT model as classifier & Achieved TTP classification recall of 0.63. \\ \hline

Rani et al.~\cite{rani2024ttpxhunter} (2024) & Techniques & CTI reports (unstructured) & Use SecureBERT for embeddings and train a linear classifier & The dataset consisted of only 193 Technique classes. Achieved a recall of 0.96 on augmented test dataset. \\ \hline

Fengrui et al.~\cite{fengrui2024few} (2024) & Techniques & MITRE Procedures (unstructured) & LLM fine-tuning with MITRE data & Achieved a recall of 0.89 when the number of samples in the fine-tuning dataset is more than 33. 
When the sample size is less, the recall achieved is 0.43. \\ \hline

Zhang et al.~\cite{zhang2024attackgboosting} (2024) & Techniques & CTI reports (unstructured) & Use LLM for similarity matching & Overall recall achieved for technique identification is 0.59. \\ \hline

Nir et al.~\cite{daniel2023labeling} (2024) & Techniques & NIDS rules (structured) & Use LLM's implicit knowledge & Maximum recall achieved with ChatGPT-4 is 0.68. \\ \hline

Mărmureanu et al.~\cite{10398612} (2023) & Tactics & SIEM rules (structured) & Train BERT model as classifier & Only 14 classes available for classification. With weight based ensemble learning strategy, achieved a recall of 0.72\\ \hline\hline

\textbf{Our method (\methodName)} & Techniques & SIEM rules (structured) & Use prompt engineering techniques and implement LLM agents to enhance LLM performance & Coverage of all technique level classes. Achieved a recall of 0.75 on the test rules. \\ \hline
\end{tabular}
\label{tab:related_work}
\end{table*}

\section{\label{sec:method}Methodology}

Each SIEM system uses its own RDL to define threat detection rules, and each RDL has its own schema.
For example, the Splunk SIEM uses the SPL to define its threat detection rules.
The task of understanding threat detection rules and recommending relevant MITRE ATT\&CK techniques (or sub-techniques) requires complex reasoning skills.
In the case of LLMs, this can be achieved with a technique called prompt chaining in which each task is divided into multiple sub-tasks in order to understand the complex reasoning behind the task.
Therefore, we employ a multi-phase architecture based on prompt chaining that leverages the power of LLMs to take a SIEM rule defined in any RDL and map it to relevant MITRE ATT\&CK techniques using the power of LLMs.
Our approach is based on the following intuitions:
\begin{itemize}[nosep,leftmargin=*]
    \item \textit{LLMs' implicit knowledge}: LLMs possess deep understanding of diverse RDLs. This enables them to interpret any rule, regardless of the RDL it is defined in, and convert it into comprehensible natural language text.
    \item \textit{LLMs' similarity comparison capability}: LLMs are adept at analyzing and comparing textual descriptions. 
    They can intelligently assess the similarity between two textual inputs to establish a meaningful connection.
\end{itemize}

\methodName has two main phases: (1) the rule to text translation phase, and (2) the MITRE ATT\&CK techniques recommendation phase.
These two phases in the pipeline include six key steps to determine relevant TTPs, as illustrated in Figure~\ref{fig:r2t}.

Although LLMs excel at translating SIEM rules into natural language, they often lack critical domain-specific contextual information related to IoCs in the rules.
To overcome this limitation, the \textit{rule to text translation} phase includes three steps: IoC extraction, contextual information retrieval, and natural language translation.

The workflow begins with the extraction of IoCs from the rules (for example, processes, log source, event codes, and file names) that the rule searches for in the logs (step (1)).In the next sstep a web search agent performs the task of obtaining additional contextual information about the IoCs discovered ((step 2)).
By incorporating this additional domain-specific information, the pipeline enhances the language translation, resulting in a more accurate and meaningful interpretation of SIEM rules.
The rule itself and the IoCs' contextual additional information from the previous stage are then used to translate the rule from RDL to natural language (step (3)).

The \textit{MITRE ATT\&CK techniques} recommendation phase of the pipeline includes the following three steps.
The rule in processed in data source identification step in which the probable origin of the data is identified (step (4)).
The description of the rule is then used to determine probable MITRE ATT\&CK techniques based on the implicit knowledge of the LLM (step (5)).
Finally, using chain-of-thought~\cite{wei2022chain} prompting, the most relevant techniques are extracted from the list of probable techniques (step (6)).
Each step of our method is further described in detail below.

\begin{figure*}[htbp]
   \includegraphics[width=\textwidth]{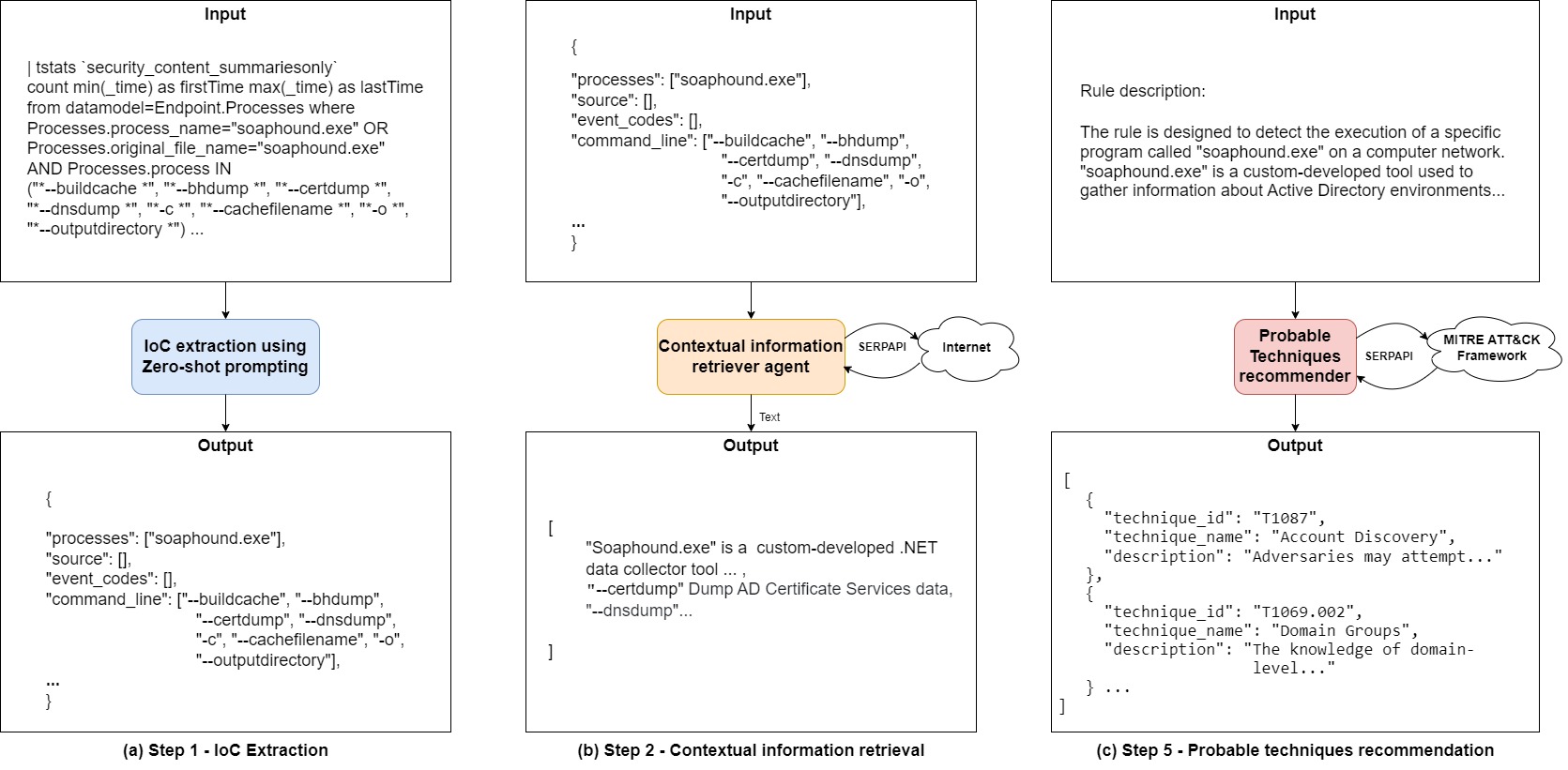}
    
   \caption{An illustration of the different steps in \methodName.}
   \label{fig:stages}
\end{figure*} 

\subsection{IoC Extraction}
The context associated with a SIEM detection rule is crucial for its accurate interpretation and effective application. 
Obtaining this contextual understanding requires comprehensive analysis of the embedded IoCs in the SIEM rule.
In the first step, \methodName systematically identifies and extracts all IoCs, identifying the types of IoCs and their corresponding values that form the foundational elements of the detection rules. 
Leveraging the LLM's inherent understanding of rule structures and IoCs, we employ a zero-shot prompting approach for this task. 
Zero-shot prompting enables the direct extraction of IoCs from the rules without requiring extensive pre-training on specific datasets.

\noindent The result of this stage is a dictionary structure, where:
\begin{itemize}[nosep,leftmargin=*]
    \item Keys represent types of IoC, such as processes, files, IP addresses, and log sources.
    \item Values are lists containing specific IoC details, such as process names, file names, IP addresses, and log source identifiers.
\end{itemize}

In the example depicted in Figure~\ref{fig:stages}(a), the pipeline processes a rule for which relevant MITRE ATT\&CK techniques need to be recommended. 
The IoC extractor LLM produces a dictionary structure as output, organizing the IoCs in a structured format to support subsequent stages in the analysis pipeline.

\subsection{Contextual Information Retrieval}
In this step, an LLM agent is employed to retrieve relevant information pertaining to the IoCs extracted from the rule.
A REACT agent~\cite{react} was used in this case to generate both reasoning traces and task-specific actions in an interleaved manner.
REACT agents interact with external tools to retrieve additional information that leads to more factual and reliable responses.
The LLM agent conducts a systematic search across web resources to gather additional contextual information for each IoC value present in the rule. 
This step addresses LLMS' lack of up-to-date knowledge or specialized domain expertise (which is critical to understanding the role and significance of the IoCs in the rule), without the need for retraining or fine-tuning.
Figure~\ref{fig:stages}(b) presents an example in which the rule includes the process name \texttt{soaphound.exe} as an IoC.
As can be seen, the web search results indicate that \texttt{soaphound.exe} is being used for active directory (AD) enumeration, which is important for the understanding of the attack. 

\subsection{Natural Language Translation}

The translation of detection rules into natural language textual descriptions fulfills three key objectives:
\begin{enumerate}[nosep,leftmargin=*]
    \item \textbf{Ensures that \methodName is format-agnostic}: It converts rules defined in various RDL formats into a generic, unstructured text format, ensuring compatibility with different SIEM systems, regardless of the specific rule format.
    \item \textbf{Provides contextual explanation}: It includes all relevant contextual information to produce a concise and comprehensible explanation of the rule.
    \item \textbf{Enhances the comprehension for LLMs}: It enables LLMs to more effectively compare the translated rule with descriptions in the MITRE ATT\&CK framework by providing a unified textual representation.
\end{enumerate}
To achieve these objectives, a zero-shot prompting technique is employed. 
The input to the LLM comprises two components:
\begin{itemize}
    \item \textbf{Syntactical information}: The rule itself, providing the structural and operational details.
    \item \textbf{Contextual information}: Details of the IoCs extracted from the rule, providing semantic insights into the rule's intent and function.
\end{itemize}
The LLM utilizes these inputs to generate a natural language textual description of the rule. 
This transformation not only ensures a more interpretable representation but also facilitates further steps of analysis and comparison, particularly in aligning the rule with MITRE ATT\&CK techniques and sub-techniques.

\subsection{Data Source or Mitigation Identification}
Identifying the most relevant data component or mitigation associated with the rule description in this step is critical for filtering out irrelevant MITRE ATT\&CK techniques (or sub-techniques) in subsequent steps of the pipeline.
In the MITRE ATT\&CK framework, data sources represent various categories of information that can be gathered from sensors or logs. 
These data sources include \textit{data components}, which are specific attributes or properties within a data source that are directly relevant to detecting a particular technique or sub-technique~. 
For example, in the context of the rule described in Figure~\ref{fig:stages}(a), the term \texttt{Endpoint.Processes} indicates that the activity is happening on an endpoint. 
Presence of the terms such as, \texttt{soaphound.exe}, \texttt{--buildcache}, \texttt{--certdump} and etc. indicate that the rule searches for command line execution of an executable named \texttt{soaphound.exe} with specific parameters. 
Therefore, the appropriate data source in this example is \textit{Command}, with the corresponding data component being \textit{Command Execution}.
Additionally, \textit{mitigations} are defined as categories of technologies or strategies that can prevent or reduce the impact of specific techniques or sub-techniques. 
The MITRE ATT\&CK framework explicitly establishes relationships between data components, mitigations, and techniques (or sub-techniques), enabling a systematic approach for identifying relevant elements.

To identify the most relevant data component or mitigation associated with a given rule description, we utilize agentic retrieval augmented generation (RAG), which incorporates an AI Agent-based implementation of the RAG framework.
Data from the MITRE ATT\&CK framework, specifically related to data components and mitigations, is stored in a vector database (e.g., ChromaDB). 
The process begins with the rule description from the previous stage, which serves as the input to the AI Agent. 
The LLM-powered agent automatically generates a search query tailored to retrieve relevant information from the RAG database.

For each query, the system retrieves the five most similar documents from the database, each containing contextual information about data components or mitigations. 
These documents are then utilized by the LLM agent to contextualize the rule description. 
By comparing the content of these retrieved documents with the rule description, the LLM agent determines and outputs the most relevant data component or mitigation along with a chain-of-thought as to why the data component or mitigation is related to the rule.

\subsection{Probable Technique Recommendation}

In this step, an LM agent is utilized to propose probable MITRE ATT\&CK techniques (and sub-techniques) that may be relevant to the description of the provided rule.
We used a REACT agent in this step as well to utilize both implicit and explicit knowledge during reasoning.
For explicit knowledge, the agent searches the MITRE ATT\&CK framework to obtain the list of probable techniques (and sub-techniques).
The natural language description of the rule from the previous step serves as input to the LLM agent.
The output of this stage consists of a list of JSON objects, each containing the MITRE technique ID, technique name, and technique description as seen in Figure~\ref{fig:stages}(c).

Throughout our experiments, we observed that as the number of recommendations ($k$) increases, both the framework's average recall and precision initially improve, however beyond a certain threshold of $k$, the 
precision begins to decline.
Based on these observations(please refer Table~\ref{tab:results3}), we selected a $k$-value of 11 to ensure a high recall.

\subsection{Relevant Technique Extraction}
In this step, \methodName refines the set of probable MITRE ATT\&CK techniques identified in the previous stage by eliminating irrelevant entries.
This step in the pipeline serves two primary purposes: (1) to enhance precision while maintaining recall achieved in previous step, and (2) to provide a clear rationale for the selection of the labels, ensuring transparency and interpretability of the mapping process.
This refinement process is grounded in the assumption that LLMs are effective for text similarity matching tasks.

The process comprises two key steps:
\begin{itemize}
    \item \textit{Rule-technique comparison}: The description of each technique in the set of probable techniques is compared with the rule description. 
    A chain-of-thought technique is then applied to elucidate the reasoning behind the association of each technique with the rule.
    \item \textit{Confidence calculation}: The generated chain-of-thought rationale for each technique (or sub-technique) is compared with the rule description to compute a relevance (or confidence) score, as done in prior work~\cite{freitas2024ai}.
\end{itemize}

Techniques with higher confidence scores are deemed more relevant to the rule. 
Conversely, techniques with scores falling below a predefined threshold are excluded.
The techniques retained after this filtering step represent the most relevant techniques corresponding to the given rule's description.

The chain-of-thought (CoT) rationale generated during the comparison of each rule to its probable technique is also provided as an output in this step.
This rationale offers a detailed natural language explanation, articulating why a particular technique is relevant to the given rule. 
Such explanations are highly valuable for security analysts, as they provide clear and transparent reasoning behind the mapping, enabling analysts to better understand and validate the association between the rule and the technique.
Other classification models proposed in previous works within this domain also suffer from the limitation of being black-box models, which lack the ability to provide clear reasoning or explanations. 
Unlike \methodName, these models fail to generate transparent, CoT rationales that explain why a particular rule is mapped to a specific technique, making them less interpretable and less useful for security analysts.
\section{Evaluation}

\subsection{Dataset}

The Splunk Security Content dataset is a comprehensive repository of security resources dedicated to improving the detection and mitigation of threats, maintained by Splunk Inc.'s research team. This dataset features over 1,600 analytic rules, all written in Splunk’s Search Processing Language (SPL), to effectively identify malicious behaviors.

These analytic rules are systematically organized into distinct domains such as endpoint, network, application, etc., based on their intended scope of applicability. For instance, the endpoint domain comprises rules specifically crafted to detect malicious activities occurring on endpoint devices. Moreover, each rule is annotated with corresponding MITRE ATT\&CK technique identifiers, thereby offering a ground-truth framework for our experiments. We chose the endpoint domain as it presents a higher diversity of MITRE ATT\&CK technique labels than other domains in the dataset.

To ensure experimental integrity, we carefully evaluated the knowledge cut-off dates of the models utilized. Among the hosted models, GPT-4-Turbo had the most recent cut-off date of December 2023,\footnote{\url{https://learn.microsoft.com/en-us/azure/ai-services/openai/concepts/models?tabs=global-standard\%2Cstandard-chat-completions\#gpt-4-and-gpt-4-turbo-models}} while Granite, the latest local model, was released in October 2024. 
To prevent any potential data leakage, we exclusively analyzed rules from the Splunk Security Content dataset that were created or modified on or after November 2024. 
During our review of the Splunk Security Content repository in December 2024, we identified 360 rules within the endpoint domain that met our criteria. 
None that none of the hosted or local models employed in our experiments were capable of inherently performing web searches since they were used via an API~\cite{chen2024}. 

\subsection{Evaluation Setup}

We implemented our method \methodName in Python 3.9, leveraging multiple libraries and frameworks for model interaction. We utilized the transformers library~\cite{transformers} for local models and for hosted models, we employed the LangChain and LangGraph frameworks.  The hosted models used in our experiments are GPT-4 Turbo, GPT-4o, and GPT-4o-mini, whereas  the local models are Mistral-7B, IBM Granite-3.0-8B, and Qwen-2.5-7B. Table~\ref{tab:models} provides a summary of all the models used in our experiments.

We configured our experimental setup with an Intel Core i7 processor, 32 GB RAM, and NVIDIA RTX 4090 GPU (24GB VRAM). Given the GPU memory constraints, we specifically selected local models with up to 8 billion parameters that could be run at full precision without quantization, ensuring optimal model performance and reliable comparison baseline. This hardware configuration allowed us to maintain consistent computational precision across all local model experiments.

With the technical infrastructure in place, we next focused on developing a standardized approach to model interaction through careful prompt engineering. Our approach follows a carefully designed template that maximizes model performance while maintaining consistency across different models. The prompts used in all steps of \methodName follow a four-component structure described below:

\begin{itemize}[nosep,leftmargin=*]
	\item \textit{Context}: This provides the background or relevant details that set the stage for the task. It includes information on the topic, scenario, or purpose, ensuring that the LLM understands the larger situation.
	\item \textit{Instruction}: This part specifies the exact task the LLM is expected to perform. It provides a clear, concise, and actionable explanation to effectively guide the LLM.
	\item \textit{Guidelines}: These are the rules or constraints that the LLM must follow when completing the task. They ensure that the output is aligned with the desired tone, format, or style.
	\item \textit{Input}: This includes any user-provided data, queries, or material required for the LLM to complete the task. It serves as the starting point or raw material for generating the output.
\end{itemize}

\begin{figure}[ht!]
    \centering
    \includegraphics[width=0.40\textwidth]{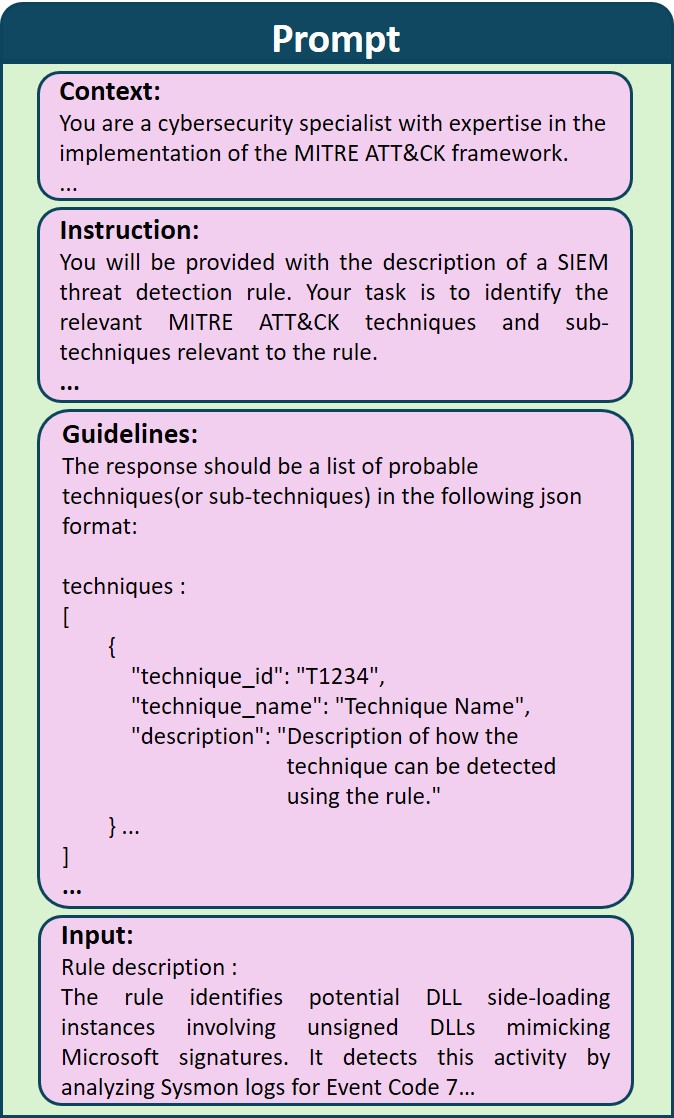}
    \caption{Overview of prompt structure used in all steps of the pipeline.}
    \label{fig:prompt}
\end{figure}

An example of a prompt used in the experiments is illustrated in Figure~\ref{fig:prompt}. 
In this example, the prompt is structured into four key components to effectively guide the LLM. 
The \textit{context} 
specifies that the task belongs to the cybersecurity domain and that the LLM should approach it as a cybersecurity specialist. 
Following this, the \textit{instruction} clearly defines the task, requiring the LLM to identify and output the relevant attack techniques/sub-techniques corresponding to the provided rule description. 
The \textit{guidelines} set additional constraints, specifying that the response must be formatted in JSON. 
Lastly, the actual rule description is presented as the \textit{input} for the LLM to process and analyze.

\begin{table*}[h]
\centering
\caption{Summary of the models and their configurations.}

\begin{adjustbox}{width=0.9\textwidth,center=\textwidth}
\begin{tabular}{|c|l|l|l|l|l|}
\hline
\multicolumn{1}{|l|}{\textbf{Model Type}} & \textbf{Model Name}  & \textbf{Parameter Size} & \textbf{Context Window} & \textbf{Max Tokens} & \textbf{Knowledge Cut-off} \\ \hline
\multirow{3}{*}{\textbf{Hosted}}          & \textbf{GPT-4-Turbo} & $\sim$1.8 Trillion      & 128,000                  & 4096                & December 2023            \\ \cline{2-6} 
         & \textbf{GPT-4o}      & $\sim$2 Trillion        & 128,000                  & 16,384               & October 2023              \\ \cline{2-6} 
           & \textbf{GPT-4o-mini} & $\sim$8 Billion         & 128,000                  & 16,384               & October 2023              \\ \hline
\multirow{3}{*}{\textbf{Local}}           & \textbf{IBM Granite} & $\sim$8 Billion         & 128,000                  & 8,192     & Not available               \\ \cline{2-6} 
             & \textbf{Mistral}     & $\sim$7 Billion         & 32,000                   & 4,096                & Not available                       \\ \cline{2-6} 
       & \textbf{Qwen}        & $\sim$3 Billion         & 128,000                  & 8,192                & Not available                       \\ \hline
\end{tabular}
\end{adjustbox}

\label{tab:models}
\end{table*}

\subsection{Baselines}

We used three baseline methods to compare \methodName with:
\begin{enumerate}[nosep,leftmargin=*]
    \item \textbf{A single GPT-4-Turbo LLM with zero-shot prompting}.
    The input to the LLM was the rule as-is, and the LLM was asked to analyze the rule and output all the MITRE ATT\&CK techniques or sub-techniques relevant to the rule.
    \item \textbf{BERT-based classifiers.} Building on the research by Mărmureanu et al.~\cite{10398612}, we trained a BERT model~\cite{devlin2018bert} as a baseline for comparison our experiments. 
    BERT's ability to capture bidirectional context through its masked language modeling (MLM) and next sentence prediction (NSP) objectives makes it particularly effective for classification tasks~\cite{sun2019fine}. 
    Additionally, we implemented CodeBERT~\cite{feng2020codebert} and an adaptation of the BERT model, as baseline models for comparison in our experiments.
    \item \textbf{TTPxHunter.} By implementing the publicly available code from the research proposed by Rani et al.~\cite{rani2024ttpxhunter}, we established an additional baseline for comparison with our method. 
    This approach utilizes SecureBERT to generate embeddings for the input data, which are then processed by a linear classifier to produce the final output.

\end{enumerate}

\subsection{Evaluation Metrics}

To evaluate our framework, we employed evaluation metrics commonly used in classic recommender systems. 
Specifically, we computed the average recall (AR) and average precision (AP) across the entire test dataset. 
In recommender systems, relevance is inherently user-specific, and both AP and AR metrics effectively capture the varying importance of individual recommendations. This characteristic makes these metrics particularly well-suited for multi-class classification problems like MITRE ATT\&CK technique prediction, where a single rule can be associated with multiple valid labels (techniques or sub-techniques).
For instance, the rule illustrated in Figure~\ref{fig:stages}(a), has multiple techniques and sub-techniques mapped to it, including: T1087.002, T1069.001, T1482, T1087.001, T1087, T1069.002, and T1069.

For computing \textit{AR}, we first calculated the recall for each sample (rule) in the dataset. 
The AR was then calculated by averaging the recall values across all samples in the dataset.
Similarly, to compute the \textit{AP}, we first calculated the precision for each sample in the dataset, and then, the AP was determined by averaging the precision values across all samples in the dataset.

In addition, the task of mapping a SIEM rule to the MITRE ATT\&CK framework is inherently a multi-class classification problem, where the number of techniques associated with each sample varies. 
Appendix~\ref{appendixA} provides an analysis of the distribution of samples based on the number of techniques they contain. 
Therefore, we also compute the Weighted Average Precision (WAP) and Weighted Average Recall (WAR) metrics. 
These metrics apply weights based on the relative number of techniques for each sample. 
\subsection{Results}

\paragraph{\textbf{Comparison of \methodName with baselines.}}
We compared \methodName with the baseline methods described in the previous section, and the results of this comparison are presented in Table~\ref{tab:Results2}. 
As evident from the table, \methodName outperformed the baseline methods when utilizing GPT-4-Turbo and GPT-4o as the underlying models. 
These findings underscore the critical role of both implicit and explicit domain-specific knowledge in enabling LLMs to deliver optimal results for the task of mapping SIEM rules to the MITRE ATT\&CK framework.

\paragraph{\textbf{\methodName's performance with different LLMs.}}
\methodName's pipeline  was executed using several LLMs to evaluate its ability to recommend MITRE ATT\&CK techniques (or sub-techniques) for a given SIEM rule. 
As can be seen in Table~\ref{tab:Results2}, GPT-4-Turbo demonstrated superior performance, delivering the most accurate and relevant recommendations when compared to other hosted and local models. 
These results provided insights into how the model configuration, including model size and architectural differences, influence the quality of recommendations generated by \methodName.

\paragraph{\textbf{Effect of model's size and context window length on performance.}}
As can be seen in Table~\ref{tab:Results2}, \methodName's performance is significantly influenced by the size of the LLMs used in its pipeline (see Table~\ref{tab:models} for different models used and their sizes and context window lengths used).
Larger models, such as GPT-4-Turbo and GPT-4o, demonstrated superior performance compared to their smaller counterparts like GPT-4o-mini or local models with fewer parameters.
This performance disparity highlights the ability of larger models to better understand complex relationships and patterns in the input data. 
The large number of parameters and greater context window allow them to capture nuanced information that smaller models might overlook, leading to more accurate and reliable recommendations.

The context window of an LLM refers to the maximum amount of text, measured in tokens, that the model can process at a time. 
This parameter plays a crucial role in determining the model's ability to handle tasks requiring extensive context.
In the case of \methodName, the performance using the Mistral model was poorer compared to other models. 
This can be attributed to the significantly smaller context window of the Mistral model, which limited its ability to process and utilize the full breadth of information required for effective recommendations.
In contrast, GPT-4o-mini, a similar model with a larger context window, performed better as it could handle more comprehensive inputs, leading to improved accuracy and reliability in mapping SIEM rules to MITRE ATT\&CK techniques

\begin{table}[h]
\centering
\caption{\methodName's performance using various hosted and local models.}
\begin{tabular}{|c|l|c|c|}
\hline
\textbf{Model Type}              & \multicolumn{1}{c|}{\textbf{Model Name}} & \textbf{AR (WAR)} & \textbf{AP (WAP)} \\ \hline
\multirow{3}{*}{\textbf{Hosted}} & \textbf{GPT-4-Turbo}                     & \textbf{0.75 (0.724)}           & \textbf{0.52 (0.51)}                \\ \cline{2-4} 
       & \textbf{GPT-4o}                          & 0.71 (0.69)                  & 0.49 (0.47)                         \\ \cline{2-4} 
         & \textbf{GPT-4o-mini}                     & 0.38 (0.36)                    & 0.29 (0.275)                        \\ \hline
\multirow{3}{*}{\textbf{Local}}  & \textbf{IBM Granite}                     & 0.34 (0.31)                    & 0.28 (0.24)                         \\ \cline{2-4} 
       & \textbf{Mistral}                         & 0.12 (0.09)                   & 0.08 (0.05)                        \\ \cline{2-4} 
    & \textbf{Qwen}                            & 0.23 (0.19)                    & 0.16 (0.15)                        \\ \hline\hline

\multirow{4}{*}{\textbf{Baselines}} & \textbf{Zero-shot LLM}                     & 0.46 (0.43)           & 0.31 (0.30)                \\ \cline{2-4} 
       & \textbf{BERT}                          &  0.68 (0.65)                   &    0.39 (0.38)                     \\ \cline{2-4} 
        & \textbf{CodeBERT}                  &  0.65 (0.63)                   &  0.47 (0.45)                      \\ \cline{2-4}
        & \textbf{TTPxHunter}~\cite{rani2024ttpxhunter}                     & 0.59 (0.56)                    & 0.42 (0.41)                        \\ \hline
\end{tabular}
\label{tab:Results2}
\end{table}

\paragraph{\textbf{Ablation study.}}
In order to analyze the importance of the different components and steps in \methodName, we performed an ablation study, mainly focusing on the impact of language translation and additional contextual information on the recommendations made by the LLM.
We performed the ablation study by running the experiment in three different scenario.
In the first scenario, the rule was provided as-is to the next stage of the pipeline; in this case, \methodName achieved an AR of approximately 0.46 and an AP of around 0.39.
In the second scenario, the rule was first translated into a natural language description, without adding any contextual information, before being passed to subsequent stages of the pipeline. 
This improved the AR to around 0.54 and the AP to approximately 0.42.
Finally, in the third scenario, the rule was translated into a natural language description enriched with contextual information before being processed by the later stages of the pipeline. 
This setup yielded the best results, with an AR of around 0.75 and an AP of approximately 0.51.
In these experiments GPT-4-Turbo was used, given its superior performance in the previous experiments performed.
A summary of the results is presented in Table \ref{tab:results1}.

These findings highlight the importance of contextual information in enhancing \methodName's performance, demonstrating that enriching natural language translations with domain-specific insights leads to improved recall and precision.

\begin{table}[h]
\centering
\caption{Impact of contextual information that enriches natural language translation with domain-specific insights on \methodName's performance.}
\begin{tabular}{|c|l|c|c|}
\hline
\textbf{S.No.} & \multicolumn{1}{c|}{\textbf{Experiment}}                                                           & \textbf{AR (WAR)} & \textbf{AP (WAP)} \\ \hline
\textbf{1}     & \textbf{Rule as-is}                                                                                & 0.46 (0.42)                   & 0.39 (0.36)                      \\ \hline
\textbf{2}     & \textbf{\begin{tabular}[c]{@{}l@{}}Rule description w/o \\ contextual   information\end{tabular}}  & 0.54 (0.49) & 0.42 (0.40)                      \\ \hline
\textbf{3}     & \textbf{\begin{tabular}[c]{@{}l@{}}Rule description with \\ contextual   information\end{tabular}} & \textbf{0.75 (0.724) }           & \textbf{0.52 (0.51) }              \\ \hline
\end{tabular}

\label{tab:results1}
\end{table}

\paragraph{\textbf{Effect of $k$ on relevant recommendations.}}

Our preliminary experiments demonstrated that limiting $k$ substantially influenced \methodName's performance. 
Larger $k$ values  improved recall by capturing a broader range of potentially relevant recommendations, but this came at the cost of reduced precision. 
This trade-off highlights the necessity of carefully selecting $k$ to balance precision and recall according to the specific needs of the application.

To address this, we replaced the hard limit on $k$ with a filtering mechanism based on the confidence (relevance) score generated in the final stage of the pipeline. Recommendations with scores below a predefined threshold were excluded. We used a threshold of 0.8, which effectively filtered low-confidence recommendations while retaining the most relevant results. This approach improved the overall performance of the model by dynamically adjusting the number of recommendations based on their relevance. \methodName's results for different $k$ values, using the abovementioned filtering mechanism, are presented in Table~\ref{tab:results3}.

\begin{table}[h]
\centering
\caption{\methodName's performance  for various values of $k$.}

\begin{tabular}{|l|l|l|l|l|}
\hline
\textbf{S.No.} & \textbf{k}          & \textbf{AR (WAR)} & \textbf{AP (WAP)} & \textbf{F1-Score} \\ \hline
\textbf{1}     & 1                                         & 0.45 (0.44) & 0.29 (0.26)                   & 0.35              \\ \hline
\textbf{2}     & 3                                         & 0.62 (0.58)   & 0.34 (0.3)                & 0.44              \\ \hline
\textbf{3}     & 5                                       & 0.66 (0.636)       & 0.45 (0.43)                & 0.53               \\ \hline
\textbf{4}     & 7                                        & 0.71 (0.68)     & 0.42 (0.40)                & 0.527               \\ \hline\textbf{5}     & 9                                       & 0.74 (0.72)     & 0.40 (0.35)           & 0.52              \\ \hline
\textbf{6}     & 11                                       & 0.75 (0.728)  & 0.39 (0.38)                   & 0.51              \\ \hline
\textbf{7}     & 13                                         & 0.75 (0.72)      & 0.35 (0.31)              & 0.48              \\ \hline
\textbf{7}     & \textbf{dynamic-k}               & \textbf{0.75 (0.724)}   & \textbf{0.52 (0.51)}         & \textbf{0.62}     \\ \hline
\end{tabular}
\label{tab:results3}
\end{table}

\begin{figure}[h!]
    \centering
    \includegraphics[width=\linewidth]{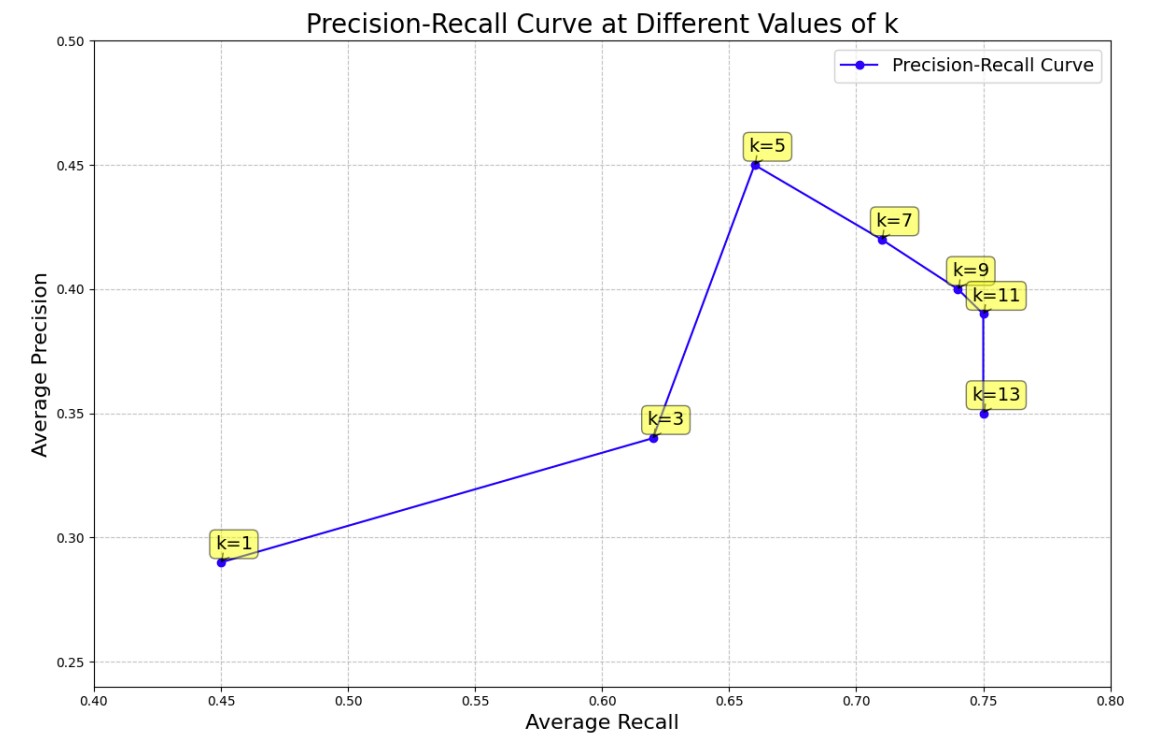}
    \caption{Average precision vs average recall curve.}
    \label{fig:pr-curve}
\end{figure}

\begin{figure}[h!]
    \centering
    \includegraphics[width=\linewidth]
    {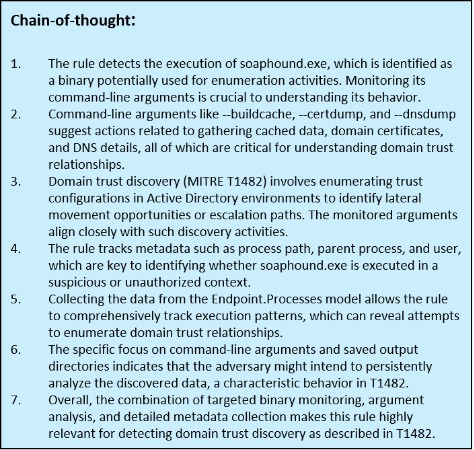}
    \caption{Chain-of-thought reasoning provided by \methodName.}
    \label{fig:cot}
\end{figure}

\paragraph{\textbf{Reasoning.}} 
One major advantage of using LLMs over other language models like BERT or CodeBERT is their inherent ability to provide reasoning for the tasks they perform. 
In \methodName's pipeline, the Rule-Technique Comparer step plays a crucial role by comparing the rule to the technique in question and generating a chain-of-thought explanation. 
This explanation, expressed in natural language, provides the relationship between the rule and the technique. 
Such transparency allows security analysts to interpret the rationale behind the mapping, enabling them to assess whether \methodName's output can be trusted or requires further scrutiny.
For example, refer to SIEM rule mentioned in the Figure~\ref{fig:stages}(a) as input.
This rule relates to the execution of an executable file called \texttt{soaphound.exe} on a system.
One of the labels for this rule is "T1482 - Domain Trust Discovery".
Figure~\ref{fig:cot} provides the CoT explanation as to why the technique T1482 is relevant to the SIEM rule.

\section{Discussion}

The use of \methodName for mapping SIEM rules to the MITRE ATT\&CK framework offers several distinct advantages. 
Unlike traditional approaches, \methodName does not require any training data, making it particularly well-suited for the cybersecurity domain, where publicly available data is often scarce. 
Additionally, the incorporation of LLMs within the pipeline enables the generation of natural language reasoning, facilitating easier interpretation of results. 
This transparency enhances the trust of security analysts on the framework, as they can better understand and validate the mapping process.

While \methodName demonstrated promising results, our study has certain limitations. 
One of the primary concerns is confidentiality, as \methodName's pipeline relies on hosted models, which may involve sending sensitive data to external servers. 
Additionally, the use of prompt chaining in the pipeline leads to longer response times compared to baseline methods. 
The duration of processing varies based on the resources available when employing local models, while for hosted models, it depends on factors such as network bandwidth and the tokens-per-second (TPS) rate of the model.

The results of our study also revealed several challenges associated with the research, which are discussed below:

\noindent \textbf{Insufficient Rule-Specific Information}: The information embedded within a rule alone was found to be inadequate for fully understanding its purpose. 
Additional contextual data such as the associated common vulnerabilities and exposures (CVE) \footnote{\url{https://cve.mitre.org/}} ID often proved necessary to interpret the rule accurately.
For example, consider the following rule:
\begin{verbatim}
| tstats `security_content_summariesonly` count FROM
datamodel=Endpoint.Registry
where Registry.registry_path="*\InProcServer32\*" 
Registry.registry_value_data=*\FORMS\* 
by Registry.registry_path Registry.registry_key_name
Registry.registry_value_name 
Registry.registry_value_data Registry.dest
Registry.process_guid Registry.user 
  | `drop_dm_object_name(Registry)` 
  | `security_content_ctime(firstTime)` 
  | `security_content_ctime(lastTime)`
\end{verbatim}
\vspace{0.2cm}
\noindent From the information provided in the rule, it is impossible to infer that this rule searches for phishing activity through \texttt{outlook.exe} process unless the CVE ID related to it is known which is "CVE-2024-21378".

\noindent \textbf{Impact of Textual Similarities on LLM Accuracy}: The high degree of similarity in textual descriptions of various MITRE ATT\&CK techniques and sub-techniques led to instances of hallucination by the language models. 
This overlap in descriptions posed a significant challenge to ensuring accurate predictions.

\noindent \textbf{Dataset Mislabeling}: The importance of a technique (or sub-technique) to a specific rule differed based on the user's perspective. 
This led to inconsistencies in how the dataset was labeled, with several cases being mislabeled. 
These errors highlight that determining relevance is often subjective.
As an example, consider the following rule,\\

\begin{verbatim}
    | tstats `security_content_summariesonly` 
    count min(_time) as firstTime max(_time) 
    as lastTime from datamodel=Change.All_Changes 
    where  All_Changes.result="*locked out*" by 
    All_Changes.user All_Changes.result 
    |`drop_dm_object_name("All_Changes")` 
    |`drop_dm_object_name("Account_Management")`
    | `security_content_ctime(firstTime)` 
    | `security_content_ctime(lastTime)` 
    | search count > 5 
\end{verbatim}
\vspace{0.2cm}
This rule aims to identify instances where excessive user account lockouts are being recorded. The label assigned to this rule in the dataset is "T1078 - Valid Accounts," which is an accurate label. 
However, the dataset does not include all the relevant labels for this rule. 
In this case, the technique "T1110 - Brute Force" is also a relevant technique that should have been mapped to the rule but was not included in the dataset.
Despite these imperfections in the labeling, the Splunk Security Content dataset provided the most reliable ground truth available for our study.

\noindent \textbf{Technique vs. Sub-Technique Prediction Challenges}: In some cases, the LLM predicted a technique instead of a sub-technique (or vice versa). 
Under stricter evaluation criteria that was used to evaluate our experiments, such discrepancies were categorized as mismatches, which negatively influenced \methodName's performance metrics of \methodName. 
These observations underline the need for more precise distinction mechanisms during prediction tasks.

\section{Conclusion and Future Work}

We proposed \methodName, an LLM-based approach for mapping SIEM threat detection rules to MITRE ATT\&CK techniques. 
While LLMs possess implicit knowledge derived from publicly available data, their direct application in cybersecurity contexts is often limited due to domain-specific challenges. 
Our experiments showed that \methodName's performance significantly improves when additional contextual information is integrated.

We identified two primary strategies for incorporating such information: (i) explicitly supplying contextual data in real time through LLM agents, and (ii) fine-tuning the LLM with domain-specific information. 
In this study, we adopted the first approach, enriching the pipeline in real time with publicly available contextual data sourced from the Internet as fine-tuning an LLM requires labelled dataset. 
As part of future work, we plan to enhance \methodName by incorporating organization-specific contextual information, which can further tailor the model to specific operational environments.
Additionally, we aim to explore fine-tuning LLMs with organization specific contextual data as an alternative approach to further improve the prediction accuracy of \methodName. 
Future research will also investigate optimization techniques such as hyperparameter tuning and ensemble methods to further enhance the performance of the proposed method. 
 With these enhancements \methodName will serve as a reliable and adaptable solution for mapping SIEM rules to MITRE ATT\&CK techniques in dynamic and complex cybersecurity landscapes.

The adoption of \methodName will contribute greatly to the automation of the cybersecurity incident response pipeline.
It also provides a roadmap for integrating advanced LLMs into the defensive strategies of organizations worldwide.

\clearpage
\bibliographystyle{ACM-Reference-Format}
\bibliography{main}

\appendix
\section{\label{appendixA}Labels Distribution}
\begin{figure}[h!]
    \centering
    \includegraphics[bb=0 0 960 480,width=1.4\linewidth]
    {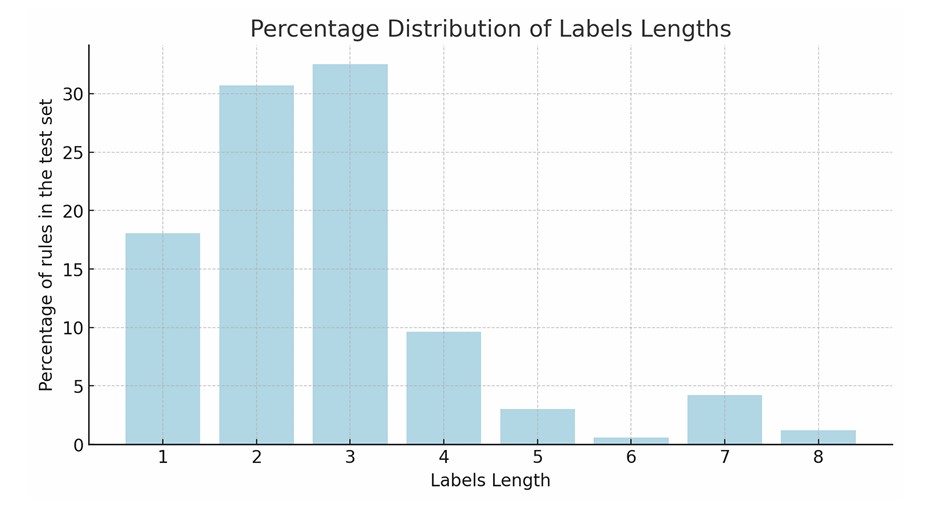}
    \caption{Distribution of length of labels in test samples.}
    \label{fig:label_distribution}
\end{figure}

\end{document}